\documentclass[12pt,preprint]{aastex}
\newcommand{\ts}{\times}

\newcommand{\beq}{\begin{equation}}
\newcommand{\bfA}{{\bf A}}

\newcommand{\bfe}{{\bf e}}
\newcommand{\curl}{{\nabla\times}}
\newcommand{\bbB}{\overline{\bf B}}
\newcommand{\bbJ}{\overline{\bf J}}
\newcommand{\bbE}{\overline{\bf E}}
\newcommand{\bbA}{\overline{\bf A}}

\newcommand{\bfa}{{\bf a}}
\newcommand{\bfB}{{\bf B}}
\newcommand{\bfb}{{\bf b}}

\newcommand{\bfu}{{\bf u}}
\newcommand{\EQ}{\begin{equation}}
\newcommand{\lb}{\langle}
\newcommand{\rb}{\rangle}
\newcommand{\EN}{\end{equation}}
\newcommand{\EQA}{\begin{eqnarray}}
\newcommand{\ENA}{\end{eqnarray}}

\newcommand{\meanBB}{\overline{\bf{B}}}

{}
{}
{}
{}
{}
{}
{}

\newcommand{\BB}{{\bf{B}}}
\newcommand{\JJ}{{\bf{J}}}

\newcommand{\AAA}{{\bf{A}}}

\newcommand{\nab}{\mbox{\boldmath $\nabla$} {}}

\newcommand{\dd}{{\rm d} {}}

\def\la{\mathrel{\mathchoice {\vcenter{\offinterlineskip\halign{\hfil
$\displaystyle##$\hfil\cr<\cr\sim\cr}}}
{\vcenter{\offinterlineskip\halign{\hfil$\textstyle##$\hfil\cr<\cr\sim\cr}}}
{\vcenter{\offinterlineskip\halign{\hfil$\scriptstyle##$\hfil\cr<\cr\sim\cr}}}
{\vcenter{\offinterlineskip\halign{\hfil$\scriptscriptstyle##$\hfil\cr<\cr\sim\cr}}}}}

\newcommand{\yjgr}[3]{ #1, {JGR,} {#2}, #3}

\newcommand{\yana}[3]{ #1, {A\&A,} {#2}, #3}

\newcommand{\ymn}[3]{ #1, {MNRAS,} {#2}, #3}

\newcommand{\ysph}[3]{ #1, {Solar Phys.,} {#2}, #3}

\newcommand{\ybook}[3]{ #1, {#2} (#3)}
\newcommand{\yproc}[5]{ #1, in {#3}, ed. #4 (#5), #2}

\begin{document}

\title{Doubly Helical Coronal Ejections from Dynamos 
and their Role in Sustaining the Solar Cycle}

\author{Eric G. Blackman$^{1}$ \& Axel Brandenburg$^{2}$}

\affil{1. Department of Physics \& Astronomy, and Laboratory for
Laser Energetics, University of Rochester, Rochester NY 14627}

\affil{2. Nordita, Blegdamsvej 17, DK-2100 Copenhagen \O, Denmark}

\begin{abstract}

Two questions about the solar magnetic field might be answered together 
once their connection is identified.
The first is important for  large scale dynamo theory:
%\cite{parker55,moffatt78,parker93}.
%$^{1-3}$
%Here, internal rotation and turbulent motions conspire to
%grow large scale magnetic structures whose direction reverses every 11 years.
%Standard models parameterize rather than solve for 
%the magnetic backreaction force on the motions driving field
%growth: 
what prevents the magnetic backreaction forces 
from shutting down the dynamo cycle?
%\cite{piddington,cattaneohughes}
%$^{4,5}$ 
The second question is: what determines the handedness of 
twist and writhe in magnetized coronal ejecta?
%\cite{rk96,chae00,demoulin02b}
%$^{6-8}$ 
Magnetic helicity conservation
%\cite{woltjer58,berger84}
%$^{9,10}$ 
is important for answering both questions. 
Conservation implies that dynamo generation of large scale writhed 
structures is accompanied by the oppositely signed twist along
these structures. The latter is   
%\cite{pfl,seehafer96,brandenburg01,fb02,paris02,brandenburg02}
%$^{11-17}$  
associated with the backreaction force. 
We suggest that coronal mass ejections (CMEs) 
%\cite{pevtsov99}
%$^{18}$
simultaneously liberate small scale 
twist and large scale writhe of opposite sign,  
helping to prevent the cycle from quenching 
and enabling a net magnetic flux change in each hemisphere.
Observations and helicity spectrum measurements from a numerical simulation of a rising 
flux ribbon in the presence of rotation support this idea.  We 
show a new pictorial of dynamo flux generation 
that includes the backreaction and magnetic helicity conservation
by characterizing the field as a 2-D ribbon rather than a 1-D line. 

\medskip

{{\bf Subject Headings}: MHD--Sun: activity--Sun: magnetic fields
MHD--turbulence; stars--magnetic fields; galaxies--magnetic fields; 
methods--numerical}

\end{abstract}

\section{Introduction}

The helical magnetic dynamo 
is the basis for a promising class of mechanisms 
to explain large scale magnetic fields observed in
stars and galaxies \cite{parker55,moffatt78,krause,parker93}.
The basic ``$\alpha-\Omega$'' dynamo is the specific version 
most relevant for strongly sheared rotators (Fig.~1). 
Interface $\alpha-\Omega$ dynamos  
(Parker 1993; Charbonneau \& MacGregor 1996; Markiel \& Thomas 1999)
include the fact that, unlike for galaxies and disks, 
the dominant shear layer is beneath the dominant turbulent region.

Focusing on the simplest ``$\alpha-\Omega$'' picture (Fig.~1), 
consider an initially weak toroidal (=encircling the rotation axis)
loop of the magnetic field embedded in the astrophysical plasma rotator.
The magnetic field is coupled to the plasma so the field 
is carried or stretched in response the plasma motion. 
Now imagine, as in the sun, 
that there is an outwardly decreasing density gradient. Consider
the motion of a rising turbulent swirl of gas, threaded by a magnetic field. 
Conservation of angular momentum dictates that the swirl 
will writhe oppositely to the underlying  rotation of the system.
This means that the initially
toroidal field threading the swirl gains a radial component.  
Statistically, rising swirls in the northern (southern) 
hemisphere writhe the field clockwise (counterclockwise). 
This is the ``$\alpha$'' 
effect and is shown by the writhed loop of Fig.~1a 
for the northern hemisphere. Differential rotation at the base of the loop
shears the radial field (the ``$\Omega$''-effect). 
The bottom part of the loop amplifies the initial toroidal 
seed loop as shown in Fig.~1b, whilst the 
top part of the loop diffused away (the ``$\beta$'' effect).  
In doing so, magnetic
flux is amplified: the flux penetrating the 
tilted rectangular surface is zero in Fig.~1a, but finite in Fig.~1b.

This process is represented mathematically 
by averaging the magnetic induction 
equation over a local volume and breaking all quantities
(velocity $\bf U$, magnetic field $\bf B$ in Alfv\'en velocity
units, and  normalized current density ${\bf J}\equiv {\curl {\bf B}}$)
into their mean (indicated by an overbar) and 
fluctuating (indicated by lower case) components. The result is 
\cite{moffatt78}:
%derived from Maxwell's equations and Ohm's law 
$\partial_t\meanBB= \curl(\alpha\meanBB+{\overline {\bf U}}\ts \meanBB) 
+(\beta+\lambda) \nabla^2\meanBB,
$
%\EQ
%\partial_t\meanBB= \curl 
%\emfb
%\meanemf +\curl({\overline {\bf V}}\ts \meanBB) + \lambda \nabla^2\meanBB,
%\label{1}
%\EN
%\beq
%\partial_t\OB=\alpha \nt \meanBB+({\beta+\lambda}) \nabla^2 \meanBB\; ,
%\label{n1}
%\ee
where $\lambda$ is the microphysical diffusivity.
%
%$\meanBB$ is the mean (large-scale) magnetic field 
%in Alfv\'en speed units and $\meanVV$ is  the mean velocity.
The ${\overline {\bf U}}$ 
term incorporates the $\Omega$-effect,
the $\beta$ term incorporates the turbulent diffusion 
(assuming constant $\beta$) 
and the first term on the right incorporates the $\alpha$-effect.
%and $\meanemf={\overline{\bfu\times\bfb}}$
%is the turbulent electromotive force, a spatially averaged correlation 
%of fluctuating velocity $\bfu$ and magnetic field $\bfb$ in Alfv\'en speed
%units.  
%Standard treatments$^{\cite{moffatt78,kr80}}$ invoke 
%$\meanemf=\alpha {\meanBB}-\beta \curl \meanBB$, 
In the kinematic theory \cite{moffatt78} 
$\alpha$ is given by $\alpha =\alpha_{0}= -(\tau/3)
{\overline {\bfu\cdot\curl\bfu}}$. 
Here $\tau$ is a 
turbulent damping time and ${\overline {\bfu\cdot\curl\bfu}}$ is the kinetic
helicity, which dictates the $\alpha$-effect described physically above. 
Usually, $\alpha$ and $\beta$ are prescribed as input parameters.

A long standing problem has been the absence of properly 
incorporating the (time-dependent) backreaction 
from the growing magnetic field on the driving turbulent motions.
%prematurely shut down the HMD [7-24]?.
%$^{4,13,16,17}$
This stimulated criticisms of mean-field  
dynamos \cite{piddington,cattaneohughes} 
and motivated interface dynamo models \cite{parker93}.
But the backreaction is now much better understood.
Steady-state studies of $\alpha$-quenching \cite{cattaneohughes}
apply only at fully saturated dynamo regimes, not at 
early times, when the backreaction is just beginning to 
be important. There the growth is fast, and most relevant for astrophysics.  
Demonstrating this  requires including the time-evolution of the 
turbulent velocity, subject to magnetic forces. 
%A self-consistent way of doing this 
%was derived in [\cite{bf02}]. 
Carrying this out formally \cite{bf02}
and using a closure in which triple correlations act as a damping
term, amounts to replacing 
$\alpha =\alpha_{0}$  by 
%$\alpha=\alpha_0+ (4\pi\tau/3c){\overline {{\bf j}\cdot\bfb}}$,
$\alpha=\alpha_0+ (\tau/3){\overline {\bfb\cdot{\curl \bfb}}}$,
where the second term is the backreaction.
It arises from $({\bf \curl \bfb}) \times \bbB$, 
the force associated with the action of the small scale current 
and the large scale field.  
This residual form of $\alpha$ has long been thought
%$^{11}$
\cite{pfl} 
to be the real driver of the helical dynamo
and has been employed in attempts to understand  
its quenching \cite{zeldovich83,kleeorin82,fb02,bb02}
%$^{15,20,21}$
%],  it became apparent that 
%a dynamical quenching model based on $\alpha_{dyn}$
%successful demonstrates 
%both a kinematic growth phase and an asymptotic resistively limited 
%phase as seen in numerical simulations [\cite{brandenburg01}] of large
%scale magnetic field growth.
From the form of $\alpha$, it is evident  
that a large current helicity 
can offset the kinetic helicity and quench the dynamo
%$^{15,20}$.
\cite{fb02,bb02}.
%This explains recent numerical simulations
%$^{14}$
%\cite{brandenburg2001}.
%We have now 
%been able to express this theory  pictorially,
%culminating in Figs.~1\&2.

In section 2 we summarize the successful backreaction theory, 
and show how it predicts ejection of twist and writhe of opposite sign. 
In section 3 we give a new pictorial 
of dynamo action that includes magnetic helicity
conservation, and discuss a simulation of a rising flux ribbon. 
Observational implications are discussed in section 4, and we conclude
in section 5. 

\section{Role of Magnetic Helicity Conservation}

The principle of magnetic helicity conservation 
determines the strength of the current helicity correction
term in $\alpha$ discussed in the previous section.
The magnetic helicity, defined by a volume integral
$H\equiv \int\AAA\cdot\BB\,\dd V\equiv\lb\AAA\cdot\BB\rb V$,
satisfies
%$^{9,10}$
\cite{woltjer58,berger84}
\begin{equation}
\partial_t H=
-2\lambda C-\mbox{surface terms},
\label{magcons}
\end{equation}
where the magnetic field 
${\bf B}$ is related to $\AAA$ by $\BB=\nab\times\AAA$, 
and the current helicity $C$ is defined by 
$C\equiv\lb{\bf J\cdot\bfB}\rb V$.
Without the surface terms (which represent
flow through boundaries)
%in astrophysical plasmas, and because $\bfJ$ typical scales
$H$ is well conserved:
for $\lambda\rightarrow 0$ the $\lambda$ term in (\ref{magcons}) 
converges to zero
%$^{22}$  
\cite{berger84b}.
% $\lambda$ 
%(as $\lambda \rightarrow 0$, ${\bf J}\propto \lambda^{-1/2}$, so
%$\lambda\lb{\bf J}\cdot{\bf B}\rb\sim\lambda^{1/2}$ converges to zero).

The magnetic helicity is  a measure of ``linkage'' and ``twist'' of field 
lines
%$^{10}$
\cite{berger84}.
%Note that upon replacement of volume integral
%by brackets, and by consider small and large scale contributions
%to the magnetic helicity, this conservation is expressed
%as $\partial_t \lb\bbA\cdot\bbB\rb + \partial_t \lb\bfa\cdot\bfb\rb 
%= \nu_M \lb{\bf J}\cdot{\bf B}\rb\simeq 0$,
Equation (\ref{magcons}) then means that in a closed 
system, the total amount of twist and writhe  is conserved.
If the large scale field is writhed one way, then the 
small scale field must twist oppositely.
In the sun, differential rotation and cyclonic
convection (the $\alpha$-effect) are both sources of 
helicity,
%$^{23}$
\cite{berger00,devore00}
but here we focus on 
the $\alpha$-effect, which generates  
large scale poloidal structures.
%demoulin 
%A rough measure of the
%relative importance of shear and helical turbulence can be obtained by
%considering the ratio of toroidal to poloidal magnetic field. For the sun
%this ratio is between 10 and 100. For poloidal magnetic field generation,
%shear does not contribute.

The $\alpha$-effect does not produce a net magnetic
twist but produces simultaneously positive
and negative magnetic twists on different scales
%$^{12-15}$
\cite{seehafer96,ji99,bf00,brandenburg2001,fb02}.
The importance of this scale segregation of $H$ 
for the backreaction term in $\alpha$ 
is  easily seen in the  two-scale approach. 
Here we write $H=H_1 + H_2$, where 
$H_1=\lb\bbA\cdot\bbB\rb V$
and $H_2=\lb{\overline {\bfa\cdot\bfb}}\rb V$ correspond to 
the volume integrated large and small scale contributions respectively.
For $C$, we then have 
$C=\lb\bbJ\cdot\bbB\rb V+\lb{\overline{{\bf j}\cdot\bfb}}\rb V
=k_1^2H_1+k_2^2H_2$,
where $k_1$ and $k_2$ represent the wavenumbers (inverse gradients) 
associated with the large and small scales respectively,
and the second equality follows rigorously for a closed system.
%(It 
%\footnote{
%can be generalized for an open system when the relative
%magnetic helicity$^8$
%$^{\cite{berger84}}$ 
%is used.) 
The current helicity backreaction in $\lb\alpha\rb$ is thus $k_2^2H_2$. 

We now relate $\bbB$  to $H_1$.
We define $\epsilon_1$ such that the large scale magnetic energy
$\lb\bbB^2\rb V= H_1{k_1/\epsilon_1}$ 
and where
${0< |\epsilon_1| \leq 1}$, where $|\epsilon_1|=1$
only for a force-free helical large scale field 
(i.e. for which $\bbJ || \bbB$ so that  the force $\bbJ \ts \bbB =0$).
In the northern hemisphere $\epsilon_1 > 0$.
By writing conservation equations analogous to (\ref{magcons}) 
for $H_1$ and $H_2$ respectively, we obtain
\begin{equation}
\partial_t H_1 = 2S
%(\lb\alpha\rb {k_1/\epsilon_1} 
%-\lb\beta \rb{k}^2_1)
%H_1
-2 \lambda k^2_1 H_1
-\mbox{surface terms} 
%\nabla\cdot\lb\rb_1,
\label{h1}
\end{equation}
%The equation for total magnetic helicity
%conservation can be written as an equation for $H_2$, namely
and 
\begin{equation}
\partial_t H_2=-2S
%( \lb\alpha \rb
%{k_1/\epsilon_1} -\lb\beta\rb k^2_1)H_1 
-2\lambda {k^2_2 H_2}
-\mbox{surface terms}
%+\nabla\cdot
%\lb\rb_2,
\label{h2}
\end{equation}
where we have used $S=(\lb\alpha\rb {k_1/\epsilon_1} 
-\lb\beta \rb{k}^2_1)H_1$ and 
$\lb\alpha\rb=
(\lb\alpha_{0}\rb + {1\over 3}\tau k^2_2
H_2/V)$.
The case without surface terms and with $\epsilon_1=1$ 
%$^{\cite{fb02,bb02}}$.
represents a dynamo without differential rotation. 
The solution \cite{fb02,bb02,bf02} 
shows that for initially  small $H_2$ but large $\alpha_{0}$,
$H_1$ grows. Growth of $H_1$ implies the oppositely signed growth of $H_2$.
This $H_2$ backreacts on $\alpha_{0}$, 
%the first terms on the right of (\ref{h1}) and (\ref{h2}), 
ultimately quenching  $\lb\alpha\rb$ and the dynamo.
%$^{\cite{brandenburg01}}$
%helical forcing at or 
%around some wavenumber $k_{\rm 2}$ leads to a bump in the 
%magnetic energy spectrum at $k<k_{\rm 2}$ 
%(larger scale) where the magnetic helicity is opposite
%in sign to that at the forcing wavenumber $k_2$. 
%As time goes on, this bump travels
%toward smaller $k$,  until it reaches the wavenumber corresponding to the 
%scale of the system $k_1$.

Since the sun is a differentially rotating open system, 
differential rotation and surface terms are important. 
The former forces $|\epsilon_{1}| < 1$, and $\epsilon_1$ 
a function of time to reflect the solar cycle.
%\footnote{By writing down the equation for magnetic energy
%associated with the large scale mean magnetic field $\lb\bbB^2\rb$
%and relating it to (\ref{h1}),  it can be shown that 
%that $k_1 <{k_1/\epsilon_1} < \Omega/\alpha_{dyn}$.)}
The presence of surface terms generally requires 
the use of the relative magnetic helicity
%$^{10}$
\cite{berger84}
%\footnote{In general, the boundary
%of the volume is not a magnetic surface, i.e.\ $\BB\cdot\nnn\neq0$, and so
%$\int\AAA\cdot\BB\,\dd V$ will not be gauge-invariant, i.e.\ the result
%will be different if one redefines $\AAA\to\AAA+\nab\phi$, where $\phi$
%is an arbitrarily chosen gauge potential. Instead one should 
%use the relative magnetic helicity$^{\cite{berger84}}$ 
%but this does not otherwise change the physical interpretations of what 
%follows.}
but to capture the key points, 
we can instead treat them as diffusion terms
%$^{24}$
\cite{bds02}.
%This means reinterpreting the volume integral represented
%by the brackets to include a large volume that includes
%protruding magnetic structures.
We combine the $\lambda$ and surface terms  
of both (\ref{h1}) and (\ref{h2}) into the forms 
 $-\nu_1 k^2_1 H_1$ and $-\nu_2 k^2_2 H_2$ respectively.
We consider the volume average  $\lb\rb$
to be taken over one hemisphere
and assume that surface terms represent diffusion into the corona.
%In the solar cycle, the magnetic field vanishes once every
%11 years. The large scale magnetic helicity in the northern
%hemisphere must oscillate from 0 to a positive maximum
%and the small scale contribution must oscillate from being
%zero to some negative minimum.  
On time scales much
shorter than the 11 year solar half cycle, 
the left sides of (\ref{h1}) and (\ref{h2}) are 
negligible and the system is in a relatively steady
state.
%\footnote{Strictly speaking, time derivatives vanish exactly 
%at solar maximum if both the large and small scale helicity
%extrema occur in phase.}
%the left sides of (\ref{h1}) and (\ref{h2}) must vanish.
%
We then see that the boundary terms are equal
and opposite
%$^{25}$
\cite{bf00}.
Here this 
implies  
%\begin{equation}
$\nu_1k_1^2 |H_1|\simeq {\nu_2 k_2^2} |H_2|$.
%\label{steady}
%\end{equation}
Since the boundary diffusion terms represent
a flux of (relative) magnetic helicity to the exterior, 
these quantities are connected to measurable observables.
{\it We therefore predict that the shedding rates 
of  small scale twist and  large scale writhe
from the $\alpha$ effect are equal in magnitude and opposite in sign.}

\section{Revising the ``Textbook'' Dynamo Pictorial}

The helicity conservation, shedding, and magnetic backreaction
are represented simply in 
Figs.~1c--e for the northern
hemisphere. The key is representing the field by 
a two-dimensional ribbon instead of a one-dimensional line.

Comparing Figs.~1a \& b ($H$ conservation not included)
with Figs.~1c \& d ($H$ conservation included) 
we see that in the latter, as 
the $\alpha$-effect produces its large scale writhe (the loop 
corresponding to gradient scale $k_1^{-1}$) 
it also twists the ribbon (corresponding to gradient scale $k_2^{-1}$). 
The large-scale writhe is right handed,
but the twist along the ribbon is left-handed and so the 
total $H$ is conserved. This is the simplest
generalization of the picture in Fig.~1a, to include 
helicity conservation. The ribbon should be thought of
as a mean field, averaged over smaller fluctuations, and 
so the actual field need not be so smooth.

The top view is shown in Fig.~1e for comparison to observations.  
In the northern
hemisphere we expect an {\sf N} shaped (or reverse-{\sf S}) sigmoid prominence
and in the southern hemisphere we would expect an {\sf S} shaped sigmoid.
In Fig.~1e we also show the backreaction force corresponding
to the small-scale magnetic twist along the ribbon: it fights against
writhing or bending the ribbon.
Eventually this twisting would suppress the $\alpha$ effect
(and thus statistically, its hemispheric average $\lb\alpha\rb$
entering (\ref{h1}) and (\ref{h2})), 
which thrives on being able to writhe the ribbon. 
In the sun, such sigmoid structures precede CMEs
%$^{14}$
\cite{pevtsov99} which on time scales of order days or weeks 
dissipate both the writhe and twist.  In doing so, they help alleviate
the backreaction on the $\alpha$ effect, and allow  a net
amplification of magnetic flux inside the sun as shown in Fig.~1c.
Some loops produced by the dynamo may not
escape, implying that some of the simultaneous 
diffusion of $H_1$ and $H_2$ is hidden in the solar interior. 
But even so, the helicity fluxes of $H_1$ and $H_2$ from the loops
which do escape, should still be equal and opposite.
Getting rid of $H_2$ simultaneously with $H_1$, either internally
or externally, is what alleviates the backreaction,
external removal by means of  CMEs is one important aspect of how this occurs.
Removal of $H_2$ implies that by solar minimum the $\alpha$ effect is again 
driven by $\alpha_0$, allowing
the cycle to repeat.

We have performed a numerical simulation 
to measure the magnetic helicity spectrum of a 
buoyant magnetic flux ribbon in the presence of rotation
which confirms the basic ideas that
twist and writhe emerge with  opposite sign.
Though rising flux tube simulations have been carried out  in the past 
%$^{26}$
\cite{abbett00}
they have not focused on the magnetic helicity spectrum. 
We started with a toroidal, horizontal flux tube 
%($y$-) 
%with vanishing
%net flux (so there is we weak oppositely oriented field outside the tube)
and a vertically dependent sinusoidal modulation of the entropy along the 
ribbon. This destabilizes the ribbon to buckle and rise in one portion.
%\footnote{The results have been obtained using the 
%{\it Pencil
%Code}, a high-order MPI code for astrophysical MHD simulations, that is
%available under {\sf http://www.nordita.dk/data/brandenb/pencil-code/}}
The boundaries were sufficiently far away to use 
a Fourier transform to obtain power spectra (Fig.~\ref{Fpspec}).
After $6$ free-fall times the spectrum shows mostly positive magnetic
helicity  together with a gradually increasing higher
wavenumber component of negative spectral helicity density. 
The latter is the anticipated contribution from 
the twist along the ribbon.

\section{Observational Implications}

In comparing the above results with observations
note that the magnetic helicity is a volume integral, so 
the ribbon on which the twisted prominence arises
may have a hidden twist elsewhere inside the sun.  
This subtlety is surmounted by use of 
the relative magnetic helicity \cite{berger84}
which allows a quantifiable interpretation of locally twisted structures.
%\footnote{This is a restatement that the 
%magnetic helicity is a gauge-dependent quantity.}
%and has to be substituted by an expression that is compatible with the
%definition of the relative magnetic helicity $^{\cite{berger84}}$
Observations typically measure the current 
helicity density, $\JJ\cdot\BB$, within a single structure, 
from which hemispheric averages can be computed, 
%or  the hemispheric averaged current helicity $\lb\JJ\cdot\BB\rb$.
%$\JJ=\nab\times\BB$ is the current density,
or the surface-integrated relative magnetic helicity fluxes. 
%of the form $\int(\EE\times\AAA)\cdot\dd {\bf S}$, wherein
%$\EE$
%=\JJ/\sigma-\uu\times\BB$ 
%is the electric field 
%$\sigma$ is the electric conductivity.

Existing observations seem to be consistent with our basic picture.
First, the observed {\sf N} sigmoids outnumber {\sf S} sigmoids
by a ratio of 6:1 in the northern  
hemisphere with the expected reverse relation in the southern hemisphere 
\cite{rk96}.  Studies of sigmoids such as Gibson et al. 2002 and refs. 
therein) do seem to show qualitative agreement with our picture.
Figure~2a of Gibson et al.\ (2002) shows a TRACE
image of an {\sf N}-sigmoid (right-handed writhe)
with left-handed twisted filament of the active
region NOAA AR 8668, typical of the northern hemisphere just
as we predict. (As our theory is statistical in nature, 
it is also not surprising that occasionally {\sf N} sigmoids 
such as AR8100 (as opposed to {\sf S} sigmoids) appear 
in the southern hemisphere \cite{green02}.  
But even for AR8100,
the  writhe is opposite in sign to the twist along the prominence.)
%This supports our prediction that magnetic helicity conservation 
%manifests to reduce the total helicity within a given 
%observable local structure of scale $k_1^{-1}$.
Second, 
%regardless of complexities, 
recent work confirms a basic hemispheric dependence of the 
sign of { small scale} current helicity,
corresponding to the twist along the ribbon in Fig.~1c.
Measurements of small scale current helicity densities 
and surface-integrated relative magnetic helicity fluxes
%$^{7,8}$
\cite{chae00,berger00}
as well as fits to line of sight magnetograms of solar
active regions
%$^{6,27-29}$
\cite{seehafer90,rk96,bao99,pevtsov00}
show primarily negative values in the north and positive in the south.
These studies measure the sign of 
%(our small scale twist), 
the twist along the ribbon not the writhe of the ribbon itself.
%whose
%sign is determined by the overall $S$ or $N$ shape of the structure discussed
%above.
We therefore agree
%$^{30}$ 
that signs of current helicity measured by these line of
sight magnetograms are characteristic of the small scale
rather than the large scale  field \cite{rs90}. 
Twist is expected at the apex of a 
writhed prominence since the density is lowest 
(Parker 1974; Choudhuri 2002).

In sum, Fig.~1e, showing an {\sf N} sigmoid is consistent
with the dominant structures of the northern hemisphere.
Large scale positive writhe dominates 
in the north, and large scale negative writhe dominates in the south.
Small scale twists along the prominences are predominately negative
in the north and positive in the south, so as  
to produce a very small net helicity in each hemisphere.
%Note again that we predict the $net$ helicity of a given 
%ejected structure to be small, since twist and
%writhe have opposite sign. 
This is  complementary to Demoulin et. al (2002) where  
oppositely signed twist an writhe from shear
were shown to be able to largely cancel, producing
a small  total magnetic helicity.  
Here we focused on the $\alpha$-effect which has the same effect.
%This is shown as the large scale poloidal field and writhe 
%in Fig.~1c, produced by the $\alpha$-effect. 
Finally note that our $\bbB$ represents 
a local averaging over the small scale twist so that 
$\bbB$ has only the writhe (Fig.~1a is thus applicable to
$\bbB$, whereas Fib 1c shows both $\bbB$ and $\bfb$).
On even larger scales, the  globally averaged field 
computed by an azimuthal average $\lb\bbB_\phi\rb$, 
is weaker than $\bbB_\phi$ in a local structure, 
due to the small filling fraction.

\section{Conclusion}

We have suggested that a new understanding of 
how helical dynamos conserve magnetic helicity  
may help resolve several mysteries of the solar magnetic field.  
We have shown that large and small scale helicities of approximately
equal magnitude should be ejected into
the solar corona as part of the sustenance of the solar cycle.
The large scale helicity  corresponds
to the writhe of a prominence, whilst the small scale helicity 
corresponds to the twist along the prominence. 
We emphasize the importance of simultaneously detecting
large and small scale contributions to the losses of
helical magnetic fields in pre-CME sigmoid structures at the solar surface.
The observations seem to be roughly consistent with our simple picture
at present but more studies and detailed modeling 
will be needed to test these ideas. Our Fig.~1 illustrates the basic concepts 
through a new pictorial representation 
of the mean field dynamo that includes magnetic helicity conservation
and the backreaction. The implications are also relevant for 
large scale dynamos in other stars and disks in astrophysics.
%The latter are also known to have explosive outflows 
%and active coronae.
%.$^{28}$
%$^{\cite{galeev79}}$

%Instead of visualizing the magnetic field strength, which can be strongly
%affected by local stretching, we visualize the rising flux tube using
%a passive scalar field that was initially concentrated along the flux
%tube. This is shown in Fig.~\ref{Fall}.

%In future simulations we plan to follow the emergence of
%the flux tube into the outer low plasma-beta exterior. We
%expect that the losses of magnetic helicity have a scale
%dependence that follows roughly that in the exterior.
%In the following subsection we discuss the consequences of surface
%losses of helical magnetic fields at small and large scales.

%\section{Conclusions}

%\acknowledgments
\noindent EB acknowledges DOE grant DE-FG02-00ER54600, and
the Dept.\ of Astrophys.\ Sciences at Princeton  for hospitality
during a sabbatical. We acknowledge  PPARC supported supercomputers at
Leicester and St Andrews, and the Odense Beowulf cluster  is acknowledged.

%\begin{references}

%\newpage

%\epsfxsize=15cm\begin{figure}[t]\epsfbox{all.eps}
\begin{figure}[t!]\plotone{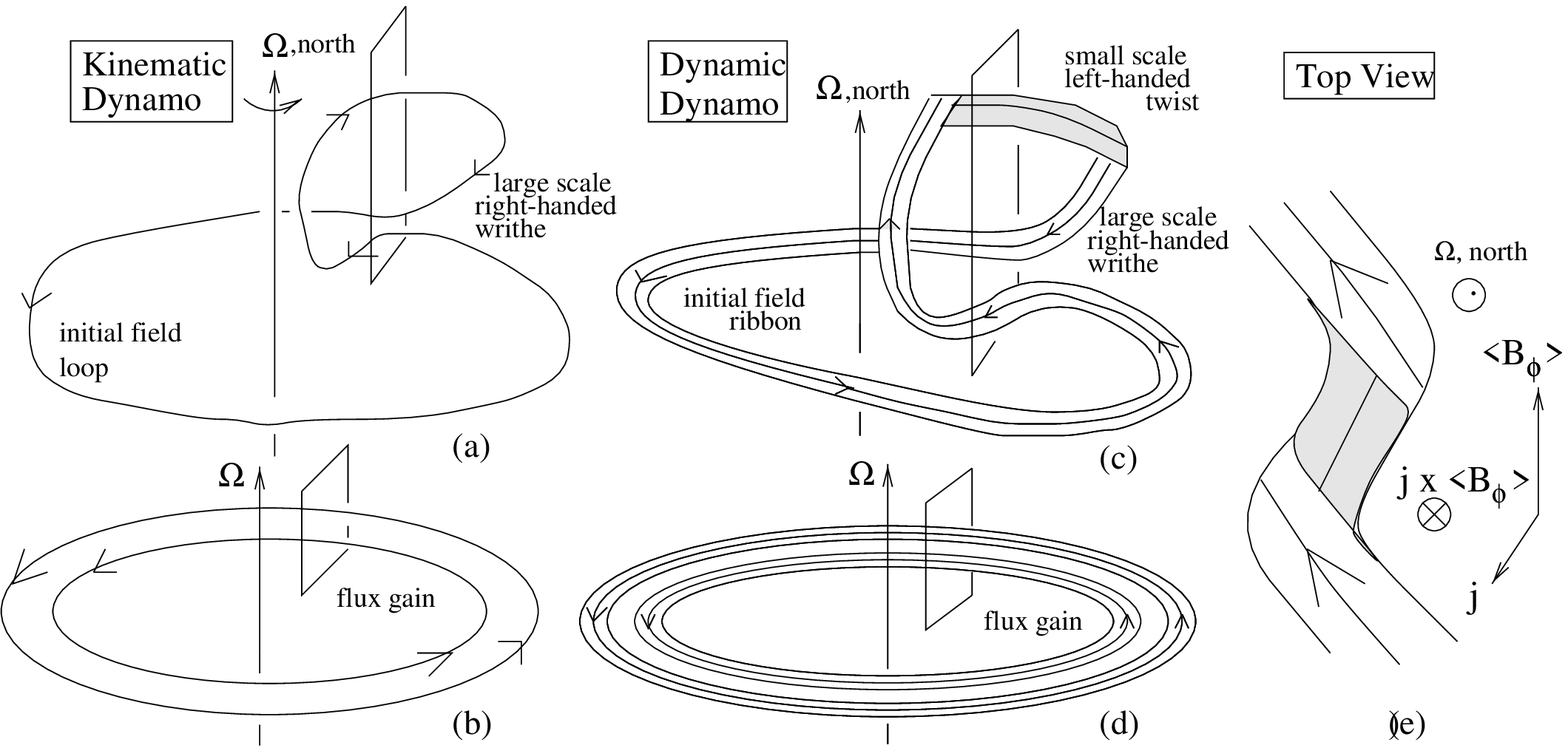}
\caption[]{ 
Schematic of {\it kinematic} helical $\alpha-\Omega$ dynamo in northern
hemisphere is shown in (a) and (b), whilst the {\it dynamic} helical 
$\alpha-\Omega$ dynamo is shown by analogy in (c) and (d).
Note that the mean field is represented as a line in (a) and (b)
and as a ribbon in (c) and (d).
(a) From an initial toroidal loop, the $\alpha$
effect induces a rising loop of right-handed writhe
that gives a radial field component.
(b) 
Differential rotation at the base of the loop 
shears the radial component, amplifying the toroidal component, 
and the ejection of the top part of loop (through coronal mass ejections) 
allow for a net flux gain
through the rectangle.  
(c) Same as (a) but now with the field represented as a flux ribbon. 
This shows how the right-handed writhe of the large scale loop 
is accompanied by a left-handed twist along the ribbon,
thus incorporating magnetic  helicity  conservation.  
(d) Same as (b) but with field represented as ribbon.
(e) Top view of the combined twist and writhe
that can be compared with observed
coronal magnetic structures in active regions. 
Note the reverse S shape of the 
right-handed large scale twist in combination with the
left-handed small scale twist along the ribbon. 
The backreaction force that resists the bending of the field
ribbon is seen to result from the small scale twist.
Note that diffusing the top part of the loops both allows for
net flux generation in the rectangles of (a)-(d), and alleviates 
the backreaction that could otherwise quench the dynamo.
}\label{all}\end{figure}

\epsscale{0.5}
\begin{figure}[t!]\plotone{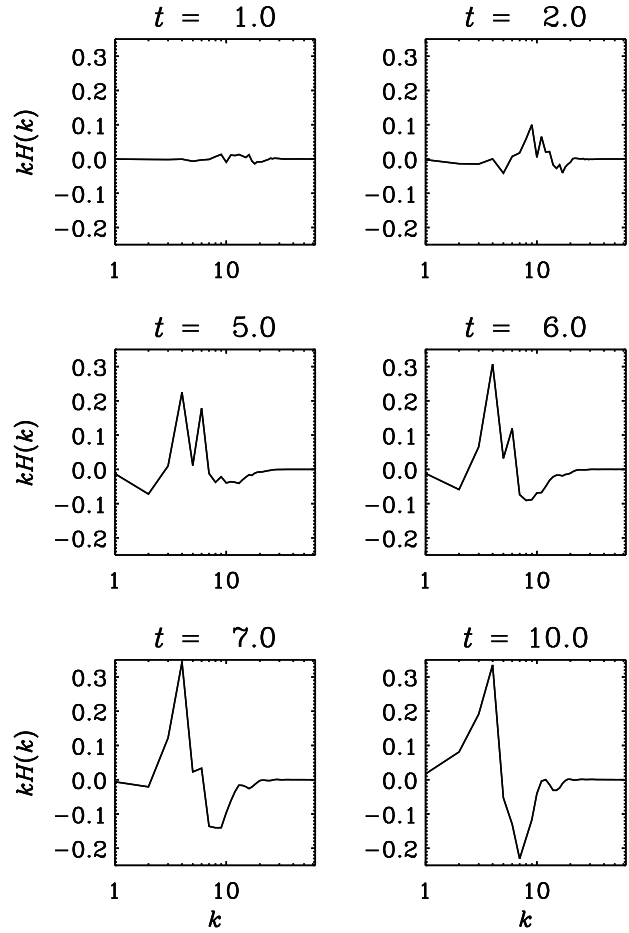}\caption[]{
Magnetic helicity spectra from rising flux ribbon simulation
(scaled by wavenumber $k$ to give magnetic
helicity per logarithmic interval) 
taken over the entire computational box. 
Positive (negative) helicity dominates at small (large) 
wavenumber.
% ($k=1-5$) 
%and negative at small scales ($k>5$).
}\label{Fpspec}\end{figure}

\end{document}